\documentclass[aps]{revtex4}
\usepackage{amsmath}
\usepackage{bm}
\usepackage{array}
\usepackage{mathrsfs}
\usepackage{epstopdf}
\usepackage{amssymb}
\usepackage{amsfonts}
\usepackage{amssymb}
\usepackage{color}

\begin{document}
\title{Local Equilibrium Spin Distribution From Detailed Balance}
\author{Ziyue Wang$^{1}$}
\author{Xingyu Guo$^{2}$}
\email{guoxy@m.scnu.edu.cn}
\author{Pengfei Zhuang$^{1}$}
\affiliation{$^{1}$ Physics Department, Tsinghua University, Beijing 100084, China\\
$^2$Guangdong Provincial Key Laboratory of Nuclear Science, Institute of Quantum Matter, South China Normal University, Guangzhou 510006, China}
\date{\today}
\begin{abstract}

As the core ingredient for spin polarization, the local equilibrium spin distribution function is derived from the detailed balance principle. The kinetic theory for interacting fermionic systems is applied to the Nambu--Jona-Lasinio model at quark level. Under the semi-classical expansion with respect to $\hbar$ and non-perturbative expansion with respect to $N_c$, the kinetic equations for the vector and axial-vector distribution functions are derived with collision terms. It is found that, for an initially unpolarized system, non-zero spin polarization can be generated at the order of $\hbar$ from the coupling between the vector and axial-vector charges. The local equilibrium spin polarization is derived from the requirement of detailed balance. It arises from the thermal vorticity and is orthogonal to the particle momentum. 

\end{abstract}
\maketitle

\section{Introduction}
\label {s1}
The single-particle distribution function is of fundamental significance in off-equilibrium kinetic theory as well as many-body physics in equilibrium state. It has long been well known that for a system consisting of identical particles in thermodynamic equilibrium, the average number of particles in a single-particle state is described by Boltzmann distribution for non-relativistic system, Bose-Einstein distribution for bosons, and Fermi-Dirac distribution for fermions. However, in general, when spin of fermions is an independent degree of freedom, the distribution for spin-$1/2$ particles needs to be extended to describe the thermodynamical equilibrium of spin degrees of freedom \cite{Becattini:2013fla}. By analyzing the density matrix for spin-$1/2$ particles, it is found that the non-even population of the polarization states arises from a steady gradient of temperature, and is orthogonal to particle momentum \cite{Becattini:2013fla}. The same equilibrium distribution is also derived in Ref.\cite{Fang:2016vpj, Gao:2018jsi} by analyzing the free streaming spin transport equation. On the experimental side, the spin polarization effect in heavy ion collisions has attracted intense attention \cite{Liang:2004ph, Voloshin:2004ha, Betz:2007kg, Becattini:2007sr}. A large global angular momentum is produced in non-central heavy ion collisions and the spin of hadrons emitted is aligned with the direction of the global angular momentum\cite{STAR:2017ckg, Adam:2018ivw, Acharya:2019vpe}. The magnitude of the global polarization of $\Lambda$ baryons can be very well described by models based on relativistic hydrodynamics and assuming local thermodynamic equilibrium of the spin degrees of freedom\cite{Becattini:2013vja, Becattini:2015ska, Becattini:2016gvu, Karpenko:2016jyx, Pang:2016igs, Xie:2017upb}. The distribution function of a system of spin $1/2$ particles is thus not only of significant importance for theoretical interest, but also required to explain the experimental data. As a matter of fact, different forms of equilibrium distribution functions are proposed based on different arguments. The most optimal situation would be to derive an equilibrium form from the entropy production\cite{Becattini:2014yxa} or the collision terms for particles with spin. As one of the basic requirements, local thermodynamic equilibrium is defined by means of detailed balance of kinetic theory, namely the collisional integral of the Boltzmann equation vanishes\cite{DeGroot:1980dk}. In this work, the equilibrium distribution functions for spin-$1/2$ particles are derived based on the detailed balance requirement.

The spin related anomalous transport phenomenon in heavy ion collisions, such as chiral magnetic effect (CME)\cite{Kharzeev:2004ey, Fukushima:2008xe} as well as chiral vortical effect\cite{Neiman:2010zi} call for the spin related transport theory and hydrodynamic theory. The chiral kinetic theory\cite{Son:2012bg, Son:2012wh, Son:2012zy, Stephanov:2012ki, Pu:2010as, Chen:2012ca, Hidaka:2016yjf, Huang:2018wdl, Liu:2018xip, Lin:2019ytz} is developed to describe the anomalous transport of massless fermions, and is further extended to the spin transport theory of massive fermions\cite{Hattori:2019ahi, Wang:2019moi, Gao:2019znl, Weickgenannt:2019dks, Liu:2020flb}. Recently, it is extended from the free streaming scenario to discussing the collisional effects\cite{Yang:2020hri, Weickgenannt:2020aaf, Carignano:2019zsh, Li:2019qkf, Hou:2020mqp}. The general framework of spin transport with collision terms is derived based on the Keldysh theory\cite{Yang:2020hri}. This framework is then applied to the weakly coupled quark-gluon plasma at high temperature to compute the spin-diffusion term for massive quarks up to the leading logarithmic order\cite{Yang:2020hri} and weakly coupled quantum electrodynamics plasma in Ref.\cite{Hou:2020mqp}. In this work, we investigate the collision term in spin transport theory based on the framework in Ref.\cite{Yang:2020hri}. In order to include fermionic 2-by-2 scattering, we consider the interaction among fermions by adopting the Nambu--Jona-Lasinio (NJL) model, and calculate the collisional self-energy by taking semi-classical ($\hbar$) expansion and non-perturbative ($1/N_c$) expansion~\cite{Klevansky:1997wm}. For massive fermions, spin is an independent degree of freedom, we take vector component $\mathcal{V}_\mu$ and axial-vector component $\mathcal{A}_\mu$ of the Wigner function as independent degrees of freedom and derive their kinetic equations at orders $\mathcal{O}(\hbar^0)$ and $\mathcal{O}(\hbar^1)$. The local equilibrium forms of vector and axial-vector components are then derived by requiring the detailed balance of the kinetic equations. Within such framework, one only needs to specify an interaction, and no more assumption is required.

The paper is organized as follows: In Section \ref{s2}, we briefly review the Wigner-function approach and derive the kinetic equations for vector and axial-vector components to the first order of $\hbar$. In Section \ref{s3}, after specifying the scalar four-fermion interaction and reviewing the free fermion solution of the classical Wigner function, we derive the local equilibrium formulae of vector and axial-vector components under the requirement of detailed balance. The spin is found to be polarized by the local vorticity. Eventually, we make concluding remarks and outlook in Section \ref{s4}. For references, we present most of the details of computations and critical steps for derivations in the Appendix.

\section{Constraint and Transport Equation}
\label {s2}
In this section, we review the basic steps of deriving the spin transport equation with collision term. Starting from the Wigner transformation applied to contour Green's function\cite{Yang:2020hri, Blaizot:2001nr}
\begin{eqnarray}
S_{\alpha\beta}^{<(>)}(X,p)=\int d^4 Y e^{ip\cdot Y/\hbar}\tilde{S}_{\alpha\beta}^{<(>)}(x,y),
\end{eqnarray}
where $X=(x+y)/2$ and $Y=x-y$ are the center of mass coordinate and relative coordinate. Here, $\tilde{S}_{\alpha\beta}^{<}(x,y)=\langle\bar{\psi}_\beta(y)\psi_\alpha(x)\rangle$ and $\tilde{S}_{\alpha\beta}^{>}(x,y)=\langle\psi_\alpha(x)\bar{\psi}_\beta(y)\rangle$ are lessor and greater propagators, respectively. The Wigner transformation of the Dyson-Schwinger equation of the lessor and greater propagators gives the Kadanoff-Baym equations\cite{Yang:2020hri}. The sum and difference of Kadanoff-Baym equations gives the constraint and transport equations. Hereafter, we focus only on $S_{\alpha\beta}^{<}(X,p)$, 
\begin{eqnarray}
\label{KBequation}
\Big\{(\gamma^\mu p_\mu-m),S^<\Big\}+\frac{i\hbar}{2}\Big[\gamma^\mu,\nabla_\mu S^<\Big]
&=&\frac{i\hbar}{2}\Big(\big[\Sigma^<,S^>\big]_\star-\big[\Sigma^>,S^<\big]_\star\Big),\nonumber\\
\Big[(\gamma^\mu p_\mu-m),S^<\Big]+\frac{i\hbar}{2}\Big\{\gamma^\mu,\nabla_\mu S^<\Big\}&=&\frac{i\hbar}{2}\Big(\{\Sigma^<,S^>\}_\star-\{\Sigma^>,S^<\}_\star\Big),
\end{eqnarray}
where $m$ is mass of the fermion, $\Sigma^{<(>)}$ are the lessor and greater self-energy. The scattering process involves only $\Sigma^{<(>)}$, thus we have dropped the real parts of the retarded and advanced self-energies and of the retarded propagators. The star product of two functions $A(q,X)$ and $B(q,X)$ is generated from the Wigner transformation, and stands for the shorthand notation of the following calculations
\begin{eqnarray}
A\star B=AB+\frac{i\hbar}{2}[AB]_{\text{P.B.}}+\mathcal{O}(\hbar^2),
\end{eqnarray}
where the Poisson bracket is $[AB]_{\text{P.B.}}\equiv(\partial_q^\mu A)(\partial_\mu B)-(\partial_\mu A)(\partial_q^\mu B)$. The commutators are $\{F,G\}\equiv FG+GF$, $[F,G]\equiv FG-GF$, $\{F,G\}_\star\equiv F\star G+G\star F$ and $[F,G]_\star\equiv F\star G-G\star F$, with $F$ and $G$ being arbitrary matrix-valued functions. 

Different Dirac components of the Wigner function have different physical meanings. Performing the spin decomposition of the Wigner function, one get various components as, 
\begin{equation}
\begin{split}
S^<=\mathcal{S}+i\mathcal{P}\gamma^5+\mathcal{V}_\mu\gamma^\mu+\mathcal{A}_\mu\gamma^5\gamma^\mu+\frac{1}{2}\mathcal{S}_{\mu\nu}\sigma^{\mu\nu},\\
S^>=\bar{\mathcal{S}}+i\bar{\mathcal{P}}\gamma^5+\bar{\mathcal{V}}_\mu\gamma^\mu+\bar{\mathcal{A}}_\mu\gamma^5\gamma^\mu+\frac{1}{2}\bar{\mathcal{S}}_{\mu\nu}\sigma^{\mu\nu},\\
\end{split}
\label{spin-propagator}
\end{equation}
where $\sigma^{\mu\nu}=i[\gamma^\mu,\gamma^\nu]/2$ and $\gamma^5=i\gamma^0\gamma^1\gamma^2\gamma^3$. 
Similarly, the collisions terms in (\ref{KBequation}) is also decomposed by the Clifford algebra, 
\begin{eqnarray}
\label{spinCD}
C=\big[\Sigma^<,S^>\big]_\star-~\big[\Sigma^>,S^<\big]_\star&=&C_S+i\gamma^5 C_P+\gamma^\mu C_{V_\mu}+\gamma^5\gamma^\mu  C_{A_\mu}+\frac{1}{2}\sigma^{\mu\nu}C_{T\mu\nu},\nonumber\\
D=\{\Sigma^<,S^>\}_\star-\{\Sigma^>,S^<\}_\star&=&D_S+i\gamma^5 D_P+\gamma^\mu D_{V_\mu}+\gamma^5\gamma^\mu  D_{A_\mu}+\frac{1}{2}\sigma^{\mu\nu}D_{T\mu\nu}.
\end{eqnarray}
Note that $C$ and $D$ contains both the loss term and the gain term, they can be recognized as $I_{\text{gain}}^c=\big[\Sigma^<,S^>\big]$, $I^c_{\text{loss}}=\big[\Sigma^>,S^<\big]$, $I_{\text{gain}}^a=\{\Sigma^<,S^>\}$ and $I_{\text{loss}}^a=\{\Sigma^>,S^<\}$, with $c$ and $a$ denoting commutator and anti-commutator respectively. Since $\Sigma$ and $S$ are both $4\times 4$ matrices, their multiplication is not commutative. 
The same spinor-basis decomposition for the self-energies is required to further derive the constraint and transport equation of each spin components, 
\begin{equation}
\begin{split}
\Sigma^<=\Sigma_S+i\Sigma_P\gamma^5+\Sigma_{V\mu}\gamma^\mu+\Sigma_{A\mu}\gamma^5\gamma^\mu+\frac{1}{2}\Sigma_{T\mu\nu}\sigma^{\mu\nu},\\
\Sigma^>=\bar{\Sigma}_S+i\bar{\Sigma}_P\gamma^5+\bar{\Sigma}_{V\mu}\gamma^\mu+\bar{\Sigma}_{A\mu}\gamma^5\gamma^\mu+\frac{1}{2}\bar{\Sigma}_{T\mu\nu}\sigma^{\mu\nu}.
\end{split}
\label{spin-Sigma}
\end{equation}

From the sum and difference of Kadanoff-Baym equations (\ref{KBequation}) as well as decomposition of the Wigner functions (\ref{spin-propagator}) and of the collision terms (\ref{spinCD}), one can derive the ten component functions
\begin{eqnarray}
\label{mastereq1}
&&p_\mu\mathcal{V}^\mu-m\mathcal{S}=\frac{i\hbar}{4}C_S,\nonumber\\
&&2m\mathcal{P}+ \hbar\nabla_\mu\mathcal{A}^\mu=-\frac{i\hbar}{2}C_P,\nonumber\\
&&2p_\mu\mathcal{S}-2m\mathcal{V}_\mu-\hbar\nabla^\nu\mathcal{S}_{\nu\mu}=\frac{i\hbar}{2}C_{V_\mu},\nonumber\\
&&\hbar\nabla_\mu\mathcal{P}-\epsilon_{\mu\nu\rho\sigma}p^\sigma\mathcal{S}^{\nu\rho}-2m\mathcal{A}^\mu=\frac{i\hbar}{2}C_{A_\mu},\nonumber\\
&&\hbar\nabla_{[\mu}\mathcal{V}_{\nu]}-2\epsilon_{\rho\sigma\mu\nu}p^\rho\mathcal{A}^\sigma-2m\mathcal{S}_{\mu\nu}=\frac{i\hbar}{2}C_{T\mu\nu},
\end{eqnarray}
and
\begin{eqnarray}
\label{mastereq2}
&&\nabla_\mu\mathcal{V}^\mu=\frac{1}{2}D_S,\nonumber\\
&&2p_\mu\mathcal{A}^\mu=\frac{\hbar}{2}D_P,\nonumber\\
&&2p^\nu\mathcal{S}_{\nu\mu}+ \hbar\nabla_\mu\mathcal{S}=\frac{\hbar}{2}D_{V_\mu},\nonumber\\
&&2p_\mu\mathcal{P}+\frac{\hbar}{2}\epsilon_{\mu\nu\rho\sigma}\nabla^\sigma\mathcal{S}^{\nu\rho}=-\frac{\hbar}{2}D_{A_\mu},\nonumber\\
&&2p_{[\mu}\mathcal{V}_{\nu]}+\hbar\epsilon_{\mu\nu\rho\sigma}\nabla^\rho\mathcal{A}^\sigma=-\frac{\hbar}{2}D_{T\mu\nu}.
\end{eqnarray}

Each component of the Wigner function and self-energies can be expanded by $\hbar$ and so as the constraint and transport equations Eq.(\ref{mastereq1}) and Eq.(\ref{mastereq2}). $\mathcal{V}$ and $\mathcal{A}$ give rise to the vector-charge and axial-charge currents through $J_V^\mu=\int q\mathcal{V}^\mu$ and $J_5^\mu=\int q\mathcal{A}^\mu$. The axial-charge currents can be regarded as a spin current of fermion. The 16 components given by the spin decomposition are not independent. Up to the first order of $\hbar$, the scalar component $\mathcal{S}$, pseudo-scalar component $\mathcal{P}$ and tensor component $\mathcal{S}_{\mu\nu}$ can be expressed in terms of $\mathcal{V}$ and $\mathcal{A}$, giving
\begin{eqnarray}
\label{Spincomponents}
\mathcal{S}^{(0)}&=&\frac{p_\mu}{m}\mathcal{V}^{(0)\mu},\nonumber\\
\mathcal{S}^{(1)}&=&\frac{p_\mu}{m}\mathcal{V}^{(1)\mu}-\frac{i}{4m}C_S^{(0)},\nonumber \\
\mathcal{P}^{(0)}&=&0,\nonumber\\
\mathcal{P}^{(1)}&=&-\frac{1}{2m}\nabla_\mu\mathcal{A}^{(0)\mu}-\frac{i}{4m}C_P^{(0)},\nonumber  \\
\mathcal{S}_{\mu\nu}^{(0)}&=&-\frac{1}{m}\epsilon_{\rho\sigma\mu\nu}p^\rho\mathcal{A}^{(0)\sigma}, \nonumber\\
\mathcal{S}^{(1)}_{\mu\nu}&=&\frac{1}{2m}\nabla_{[\mu}\mathcal{V}^{(0)}_{\nu]}-\frac{1}{m}\epsilon_{\rho\sigma\mu\nu}p^\rho\mathcal{A}^{(1)\sigma}-\frac{i}{4m}C^{(0)}_{T\mu\nu}.
\end{eqnarray}
Each of $\mathcal{V}_\mu$ and $\mathcal{A}_\mu$ contains 4 components, which are not all independent. $p_\mu\mathcal{A}^{(0)\mu}=0$ and $p_{[\mu}\mathcal{V}^{(0)}_{\nu]}=0$ indicates that $\mathcal{A}^{(0)}_\mu$ has three independent components, while $\mathcal{V}^{(0)}_{\mu}$ has only one independent component. Because of similar restrictions at $\mathcal{O}(\hbar)$, the number of independent components of $\mathcal{A}^{(1)}_\mu$ and $\mathcal{V}^{(1)}_{\mu}$ stays the same as at $\mathcal{O}(\hbar^0)$. In order to keep the description covariant and symmetric, we in the following derive the transport equations of $\mathcal{V}_\mu$ and $\mathcal{A}_\mu$, but keep in mind that $\mathcal{V}_\mu$ and $\mathcal{A}_\mu$ has redundant components, and that the system has 4 independent degrees of freedom in total, one for number density and three for spin density.

The classical components are on the mass shell $(p^2-m^2)\mathcal{V}^{(0)}_\mu=0$, $(p^2-m^2)\mathcal{A}^{(0)}_\mu=0$. The transport equations are 
\begin{eqnarray}
&& (p\cdot\nabla)\mathcal{V}^{(0)}_\mu=\frac{m}{2}D_{V\mu}^{(0)}+\frac{i}{2}p^\nu C_{T\nu\mu}^{(0)},\nonumber\\
&&(p\cdot\nabla)\mathcal{A}^{(0)}_\mu=\frac{m}{2}D_{A\mu}^{(0)}-\frac{i}{2}p_\mu C_P^{(0)},
\end{eqnarray}
With the spin decomposition of collision terms $C$ and $D$ given in Appendix.\ref{a1}, the transport equations of the vector and axial-vector components become, 
\begin{eqnarray}
p\cdot\nabla\mathcal{V}^{(0)}_\mu
&=&m\widehat{\Sigma_S^{(0)}\mathcal{V}_{\mu}^{(0)}}
+p^\nu\widehat{\Sigma_{V\nu}^{(0)}\mathcal{V}_{\mu}^{(0)}}+\frac{m}{2}\epsilon_{\alpha\beta\lambda\mu}\widehat{\Sigma_{T}^{(0)\alpha\beta}\mathcal{A}^{(0)\lambda}}-\frac{p_\nu}{m}\epsilon_{\alpha\mu\beta\lambda}p^\beta\widehat{\Sigma_T^{(0)\alpha\nu}\mathcal{A}^{(0)\lambda}}-p_\mu\widehat{\Sigma_{A}^{(0)\nu}\mathcal{A}_{\nu}^{(0)}},\nonumber\\
\label{TA0}
p\cdot\nabla\mathcal{A}^{(0)}_\mu
&=&m\widehat{\Sigma_S^{(0)}\mathcal{A}_\mu^{(0)}}
+p^\nu\widehat{\Sigma_{V\nu}^{(0)}\mathcal{A}_\mu^{(0)}}
+\frac{m}{2}\epsilon_{\alpha\beta\lambda\mu}\widehat{\Sigma_T^{(0)\alpha\beta}\mathcal{V}^{(0)\lambda}}
+\widehat{\Sigma_{A\mu}^{(0)}p^\nu\mathcal{V}_\nu^{(0)}}
-p_\mu\widehat{\Sigma_{A\nu}^{(0)}\mathcal{V}^{(0)\nu}},
\end{eqnarray}
where, same as in Ref.\cite{Yang:2020hri}, the hat operator is defined as $\widehat{FG}=\bar{F}G-F\bar{G}$. Since the spin polarization is in general a quantum effect, it is crucial to investigate the transport equation at the first order of $\hbar$, especially the transport equation of $\mathcal{A}_\mu^{(1)}$. Taking the semiclassical expansion of Eq.(\ref{mastereq1}) and Eq.(\ref{mastereq2}), and considering the relation between various spin components Eq.(\ref{Spincomponents}), the on-shell conditions and transport equations at $\mathcal{O}(\hbar)$ are modified into
\begin{eqnarray}
&&(p^2-m^2)\mathcal{V}^{(1)}_\mu=\frac{ip_\mu}{4}C_S^{(0)}+\frac{im}{4}C_{V\mu}^{(0)},\nonumber\\
&&(p^2-m^2)\mathcal{A}_\mu^{(1)}=\frac{1}{4}p^\mu D_P^{(0)}-\frac{i}{8}\epsilon_{\mu\alpha\beta\gamma}p^\alpha C_{T}^{(0)\beta\gamma}+\frac{i}{4}mC_{A\mu}^{(0)},\nonumber\\
&&(p\cdot\nabla)\mathcal{V}_\mu^{(1)}=\frac{m}{2}D_{V\mu}^{(1)}+\frac{i}{2}p^\nu C_{T\nu\mu}^{(1)}+\frac{i}{4}\nabla_\mu C_S^{(0)},\nonumber\\
&&(p\cdot\nabla)\mathcal{A}^{(1)}_\mu=\frac{m}{2}D_{A\mu}^{(1)}-\frac{i}{2}p_\mu C_P^{(1)}-\frac{i}{8}\epsilon_{\mu\sigma\nu\rho}\nabla^\sigma C_{T}^{(0)\nu\rho},\end{eqnarray}
together with the restrictions at first order of $\hbar$, $p_\mu\mathcal{A}^{(1)\mu}=\frac{1}{4}D_P^{(0)}$ and $p_{[\mu}\mathcal{V}^{(1)}_{\nu]}=-\frac{1}{2}\epsilon_{\mu\nu\rho\sigma}\nabla^\rho\mathcal{A}^{(0)\sigma}-\frac{1}{4}D^{(0)}_{T\mu\nu}$. With the spin decomposition and semi-classical expansion of collision terms $C$ and $D$ in Appendix.\ref{a1}, as well as the relation between the spin components of the Wigner function Eq.(\ref{Spincomponents}), the on-shell relation become
\begin{eqnarray}
(p^2-m^2)\mathcal{V}^{(1)}_\mu
&=&-\frac{m}{2}\widehat{\Sigma_P^{(0)}\mathcal{A}_\mu^{(0)}}
-\frac{1}{2}\epsilon_{\mu\nu\alpha\beta}p^\alpha\widehat{\Sigma_{V}^{(0)\nu}\mathcal{A}^{(0)\beta}}
-\frac{m}{2}\widehat{\Sigma_{T\mu\nu}^{(0)}\mathcal{V}^{(0)\nu}},\nonumber\\
(p^2-m^2)\mathcal{A}_\mu^{(1)}
&=&\frac{p^\mu}{2m}p^\nu\widehat{\Sigma_P^{(0)}\mathcal{V}_\nu^{(0)}}-\frac{m}{2}\widehat{\Sigma_P^{(0)}\mathcal{V}_\mu^{(0)}}+\epsilon_{\mu\alpha\beta\gamma}p^\alpha\widehat{\Sigma_{A}^{(0)\beta}\mathcal{A}^{(0)\gamma}},
\end{eqnarray}
In general, at $\mathcal{O}(\hbar)$ order,  $\mathcal{V}^{(1)}_\mu$ and $\mathcal{A}^{(1)}_\mu$ are off-shell because of the interaction. The restrictions at $\mathcal{O}(\hbar)$ again eliminate the redundant components in $\mathcal{A}^{(1)}_\mu$ and $\mathcal{V}^{(1)}_\mu$,
\begin{eqnarray}
p_{[\mu}\mathcal{V}^{(1)}_{\nu]}
&=&-\frac{1}{2}\epsilon_{\mu\nu\alpha\beta}\nabla^\alpha\mathcal{A}^{(0)\beta}
+\frac{1}{2m}\epsilon_{\mu\nu\alpha\beta}p^\alpha\widehat{\Sigma_S^{(0)}\mathcal{A}^{(0)\beta}}
+\frac{1}{2}\epsilon_{\mu\nu\alpha\beta}\widehat{\Sigma_{V}^{(0)\alpha}\mathcal{A}^{(0)\beta}}+\frac{1}{2m}p_{[\mu}\widehat{\Sigma_P^{(0)}\mathcal{A}_{\nu]}^{(0)}}\nonumber\\
&&-\frac{1}{2m}p^\rho\widehat{\Sigma_{T\mu\nu}^{(0)}\mathcal{V}_\rho^{(0)}}
-\frac{1}{2}\epsilon_{\mu\nu\alpha\beta}\widehat{\Sigma_{A}^{(0)\alpha}\mathcal{V}^{(0)\beta}},\nonumber\\
p_\mu\mathcal{A}^{(1)\mu}&=&\frac{1}{2m}p^\nu\widehat{\Sigma_P^{(0)}\mathcal{V}_\nu^{(0)}}+\frac{1}{2m}p^\rho\widehat{\Sigma_{T\rho\nu}^{(0)}\mathcal{A}^{(0)\nu}}.
\end{eqnarray}
Since the RHS of the restrictions contains only the $\mathcal{O}(\hbar^0)$ components, $\mathcal{V}^{(1)}_\mu$ contains still only one independent component representing first order correction to number density, while $\mathcal{A}^{(1)}_\mu$ contains three independent components representing first order correction to spin density.

The transport equations of the first order components $\mathcal{V}^{(1)}_\mu$ and $\mathcal{A}^{(1)}_\mu$ are
\begin{eqnarray}
\label{TV1}
(p\cdot\nabla)\mathcal{V}^{(1)}_\mu&=&
+m\widehat{\Sigma_S^{(0)}\mathcal{V}_{\mu}^{(1)}}
+p^\nu\widehat{\Sigma_{V\nu}^{(0)}\mathcal{V}_{\mu}^{(1)}}
-p_\mu\widehat{\Sigma_{A}^{(0)\nu} \mathcal{A}^{(1)}_{\nu}}
-\frac{p_\nu}{m}\epsilon_{\rho\sigma\alpha\mu}p^\rho\widehat{\Sigma_{T}^{(0)\alpha\nu}\mathcal{A}^{(1)\sigma}}
+\frac{m}{2}\epsilon_{\sigma\nu\lambda\mu}\widehat{\Sigma_{T}^{(0)\sigma\nu}\mathcal{A}^{(1)\lambda}}\\
&&
+m\widehat{\Sigma_S^{(1)}\mathcal{V}_{\mu}^{(0)}}
+p^\nu\widehat{\Sigma_{V\nu}^{(1)}\mathcal{V}_{\mu}^{(0)}}
-p_\mu\widehat{\Sigma_{A}^{(1)\nu}\mathcal{A}_\nu^{(0)}}
-\frac{p_\nu}{m}\epsilon_{\alpha\mu\beta\lambda}p^\beta\widehat{\Sigma_{T}^{(1)\alpha\nu}\mathcal{A}^{(0)\lambda}}+\frac{m}{2}\epsilon_{\sigma\nu\lambda\mu}\widehat{\Sigma_{T}^{(1)\sigma\nu}\mathcal{A}^{(0)\lambda}}\nonumber\\
&&
+\frac{1}{2m}p^\nu [\widehat{\Sigma_{T\mu\nu}^{(0)}(p^\alpha\mathcal{V}_\alpha^{(0)}})]_{\text{P.B.}}
-\frac{m}{2}[\widehat{\Sigma_{T\mu\nu}^{(0)}\mathcal{V}^{(0)\nu}}]_{\text{P.B.}}
+\frac{1}{2}\epsilon_{\mu\nu\alpha\beta}p^\nu [\widehat{\Sigma_{A}^{(0)\alpha}\mathcal{V}^{(0)\beta}}]_{\text{P.B.}}\nonumber\\
&&
-\frac{1}{2m}p_\nu\widehat{\Sigma_{T\alpha\mu}^{(0)}\nabla^{[\alpha}\mathcal{V}^{(0)\nu]}}
+\frac{1}{2m}p_\nu\widehat{\Sigma_{T}^{\alpha\nu(0)}\nabla_{[\alpha}\mathcal{V}^{(0)}_{\mu]}}
+\frac{1}{2}\epsilon_{\beta\nu\lambda\mu}\widehat{\Sigma_{A}^{(0)\beta}\nabla^{\nu}\mathcal{V}^{(0)\lambda}}
\nonumber\\
&&
+\frac{1}{2}\epsilon_{\mu\nu\alpha\beta}(\nabla^\alpha\widehat{\Sigma_{V}^{(0)\nu})\mathcal{A}^{(0)\beta}}
-\frac{1}{2m}p_\mu(\nabla^\nu\widehat{\Sigma_P^{(0)})\mathcal{A}_\nu^{(0)}}
-\frac{1}{2m}(p^\nu\nabla_\nu\widehat{\Sigma_P^{(0)})\mathcal{A}_\mu^{(0)}}
+\frac{1}{2m}p^\nu\epsilon_{\mu\nu\alpha\beta}(\nabla^\alpha\widehat{\Sigma_S^{(0)})\mathcal{A}^{(0)\beta}},\nonumber
\end{eqnarray}
and
\begin{eqnarray}
\label{TA1}
(p\cdot\nabla)\mathcal{A}_\mu^{(1)}
&=&
+m\widehat{\Sigma_S^{(0)}\mathcal{A}^{(1)}_{\mu}}
+p^\nu\widehat{\Sigma_{V\nu}^{(0)}\mathcal{A}^{(1)}_{\mu}}
+p^\nu\widehat{\Sigma_{A\mu}^{(0)}\mathcal{V}^{(1)}_{\nu}}
+\frac{m}{2}\epsilon_{\alpha\beta\lambda\mu}\widehat{\Sigma_T^{(0)\alpha\beta}\mathcal{V}^{(1)\lambda}}
-p_\mu\widehat{\Sigma_{A\nu}^{(0)}\mathcal{V}^{(1)\nu}}\\
&&
+m\widehat{\Sigma_S^{(1)}\mathcal{A}_\mu^{(0)}}
+p^\nu\widehat{\Sigma_{V\nu}^{(1)}\mathcal{A}_\mu^{(0)}}
+p^\nu\widehat{\Sigma_{A\mu}^{(1)}\mathcal{V}_\nu^{(0)}}
+\frac{m}{2}\epsilon_{\alpha\beta\lambda\mu}\widehat{\Sigma_T^{(1)\alpha\beta}\mathcal{V}^{(0)\lambda}}
-p_\mu\widehat{\Sigma_{A\nu}^{(1)}\mathcal{V}^{(0)\nu}}\nonumber\\
&&
-\frac{1}{2}\epsilon_{\mu\nu\rho\sigma}(\nabla^\sigma\widehat{\Sigma_{V}^{(0)\nu})\mathcal{V}^{(0)\rho}}
-\frac{m}{2}[\widehat{\Sigma_P^{(0)}\mathcal{V}_\mu^{(0)}}]_{\text{P.B.}}
+\frac{1}{2m}p_\mu[\widehat{\Sigma_P^{(0)}(p^\nu\mathcal{V}_\nu^{(0)}})]_{\text{P.B.}}\nonumber\\
&&
+\frac{1}{2}\epsilon_{\mu\sigma\nu\rho}\nabla^\sigma\widehat{\Sigma_{A}^{(0)\nu}\mathcal{A}^{(0)\rho}}
+\frac{1}{2}\epsilon_{\nu\mu\alpha\beta}[\widehat{\Sigma_{A}^{(0)\nu}(p^\alpha\mathcal{A}^{(0)\beta}})]_{\text{P.B.}}
-\frac{m}{2}[\widehat{\Sigma_{T\mu\nu}^{(0)}\mathcal{A}^{(0)\nu}}]_{\text{P.B.}}
+\frac{1}{2m}p_\mu[\widehat{\Sigma_{T\rho\nu}^{(0)}(p^\rho\mathcal{A}^{(0)\nu}})]_{\text{P.B.}}\nonumber\\
&&
-\frac{1}{2m}p_\sigma\nabla^\sigma(\widehat{\Sigma_{T\mu\nu}^{(0)}\mathcal{A}^{(0)\nu}})
+\frac{1}{2m}p^\nu\nabla^\sigma(\widehat{\Sigma_{T\mu\nu}^{(0)}\mathcal{A}_\sigma^{(0)}})
+\frac{1}{2m}p_\mu\nabla^\sigma(\widehat{\Sigma_{T\sigma\nu}^{(0)}\mathcal{A}^{(0)\nu}})
-\frac{1}{2m}p^\nu\nabla^\sigma(\widehat{\Sigma_{T\sigma\nu}^{(0)}\mathcal{A}_\mu^{(0)}}).\nonumber
\end{eqnarray}
The first two lines in Eq.(\ref{TV1}) and Eq.(\ref{TA1}) are dynamical effects, which contain for instance the diffusion effect. These terms have the same structure as the collision terms in the classical limit (\ref{TA0}). The last three lines in both transport equations relate to the derivatives of self-energies and distribution functions, which are inhomogeneous effects. These terms are also quantum effects, which generate the coupled transport of vector charge and axial-vector charge. As we will see in the following, these inhomogeneous effects produce spin polarization from the thermal vorticity. 

\section{Fermionic $2$ by $2$ scattering}
\label{s3}
In this paper, we focus on deriving the local equilibrium distribution from the detailed balance principle. For this purpose, the interaction needs to be specified to calculate the explicit expression of off-diagonal self-energies $\Sigma^<$ and $\Sigma^>$. Considering the fact that, different interaction determines only how fast the system reaches equilibrium state, but not the equilibrium distribution function, therefore, we adopt the NJL-type model with scalar-channel of interaction and calculate the fermionic $2$ by $2$ scattering,
\begin{eqnarray}
\mathcal{L}=\bar\psi(i\hbar \partial\!\!\!/-m)\psi+G(\bar\psi\psi)^2.
\end{eqnarray}
In general a large part of the light fermion mass comes from the chiral condensate, however, we here work in the chiral restored phase to simplify the calculation, and consider only the current mass. Due to the nature of the strong coupling theory, two expansions must be applied: one expansion in the inverse number of colors $1/N_c$, and one semiclassical expansion in powers of $\hbar$. Directly translating from the diagrams\cite{Klevansky:1997wm}, and then perform the Wigner transformation, this results in the self-energy to the $1/N_c$ order (denoted by LO) and to the $1/N_c^2$ order (denoted by NL)
\begin{eqnarray}
\label{SigmaLO}
\Sigma^{>}_{\text{LO}}(X,p)&=&G^2\int dP~S^{>}(X,p_1)\text{Tr}\Big[S^{<}(X,p_2)S^{>}(X,p_3)\Big],\nonumber\\
\Sigma^{>}_{\text{NL}}(X,p)&=&-G^2\int dP~S^{>}(X,p_1)S^{<}(X,p_2)S^{>}(X,p_3),
\end{eqnarray}
with the momentum integral defined as $\int dP=\int\frac{d^4p_1d^4p_2d^4p_3}{(2\pi)^4(2\pi)^4(2\pi)^4}(2\pi )^4\delta(p-p_1+p_2-p_3)$. The lesser self-energy $\Sigma^{<}_{\text{LO}}$ and $\Sigma^{<}_{\text{NL}}$ can be obtained by taking the exchange $S^>\leftrightarrow S^<$ from (\ref{SigmaLO}). As clarified in \cite{Klevansky:1997wm}, the self-energy $\Sigma^{<(>)}_{\text{LO}}$ and $\Sigma^{<(>)}_{\text{NL}}$ correspond to different scattering channels. However, the spin decomposition of $\Sigma^{<(>)}_{\text{NL}}$ is much more complicated when the Green's function involves Dirac structure. Besides, since the detailed balance requires that the gain term and the loss term cancel with each other in arbitrary collision channel, to simplify the calculation, we consider only the collisional self-energy at $\mathcal{O}(1/N_c)$ order. The spin decomposition of $\Sigma_{\text{LO}}$ follows simply from that of $S^{<(>)}(X,p_1)$, since the $\text{Tr}(S^<(X,p_2)S^>(X,p_3))=\mathcal{S}^2\bar{\mathcal{S}}^3-\mathcal{P}^2\bar{\mathcal{P}}^3+\mathcal{V}^2_\mu\bar{\mathcal{V}}^{3\mu}-\mathcal{A}^2_\mu\bar{\mathcal{A}}^{3\mu}+\frac{1}{2}\mathcal{S}^2_{\mu\nu}\bar{\mathcal{S}}^{3\mu\nu}$ is a number.
The self-energy $\Sigma^>_{\text{LO}}$ can be decomposed as $\bar\Sigma^{\text{LO}}_i=G^2\int dP~\text{Tr}\big(S^{<}(X,p_2)S^{>}(X,p_3)\big){S}^<_i(X,p_1)$. For instance, $\bar{\Sigma}_{S}$ corresponds to $\bar{\mathcal{S}}$, $\bar{\Sigma}_{V\mu}$ corresponds to $\bar{\mathcal{V}}_\mu$, and $\bar{\Sigma}_{A\mu}$ corresponds to $\bar{\mathcal{A}}_\mu$ and so on. 

\subsection{Classical Limit}
Substituting in the spin components of the self-energy in the transport equation of $\mathcal{V}^{(0)}_\mu$ and $\mathcal{A} ^{(0)}_\mu$ in Eq.(\ref{TA0}), one has the transport equations including the collision terms. Considering the relation between $\mathcal{S}^{(0)}$ and $\mathcal{V}^{(0)}_\mu$, and that $\mathcal{V}^{(0)}_\mu$ can be decomposed to $\mathcal{V}^{(0)}_\mu\propto\delta(p^2-m^2)p^\mu f_V$, it would be convenient to derive the collision terms in transport equation of vector charge distribution $f_V$ and the axial-vector charge distribution $f_A$ from the following two equations, 
\begin{eqnarray}
\label{TSLO}
p\cdot\nabla \mathcal{S}^{(0)}
&=&G^2\int dP\Big[\Big(1+\frac{p_2\cdot p_3}{m^2}\Big)\left(\widehat{\mathcal{S}^2{\mathcal{S}}^3}-\widehat{\mathcal{A}^{2\mu}\mathcal{A}^{3}_\mu}\right)+\frac{p_3^\mu p_2^\nu}{m^2}\widehat{{\mathcal{A}^2_\mu\mathcal{A}}^{3}_\nu}\Big]\Big[\Big(m+\frac{p\cdot p_1}{m}\Big)\left(\widehat{\mathcal{S}{\mathcal{S}}^1}- \widehat{\mathcal{A}^\nu{\mathcal{A}}^{1}_\nu}\right)
+\frac{p_{1}^\mu p^\nu}{m} \widehat{\mathcal{A}^\mu{\mathcal{A}}^{1}_\nu}\Big],\nonumber\\
\label{TALO}
p\cdot\nabla\mathcal{A}^{(0)}_\mu
&=&G^2\int dP\Big[\Big(1+\frac{p_2\cdot p_3}{m^2}\Big)\left(\widehat{\mathcal{S}^2{\mathcal{S}}^3}-\widehat{\mathcal{A}^{2\mu}\mathcal{A}^{3}_\mu}\right)+\frac{p_3^\mu p_2^\nu}{m^2}\widehat{{\mathcal{A}^2_\mu\mathcal{A}}^{3}_\nu}\Big]\Big[\Big(m+\frac{p\cdot p_1}{m}\Big)(\widehat{\mathcal{A}_\mu{\mathcal{S}}^1}
+\widehat{\mathcal{S}\mathcal{A}_\mu^1})
-\frac{p_\mu+p_{1\mu}}{m}
p^\nu\widehat{\mathcal{S}\mathcal{A}^1_\nu}\Big].\nonumber\\
\end{eqnarray}
Note that all the components on the right hand side are at leading order of $\hbar$. Before moving on to analyzing the scattering channels, we first recall the classical free fermion solution of the various components of Wigner function \cite{Weickgenannt:2019dks}. From the definition of the Wigner function as well as the contour green's function, the classical Wigner function in a free fermion system is given by 
\begin{eqnarray}
S^<(X,p)&=&\frac{1}{(2\pi)^3}\delta(p^2-m^2)\sum_{sr}\Big\{\theta(p^0)\bar{u}_s(\mathbf{p})
{u}_r(\mathbf{p})f_q^{sr}(X,p)+\theta(-p^0)\bar{v}_s(-\mathbf{p})
{v}_r(-\mathbf{p})\bar{f}_{\bar{q}}^{sr}(X,-p)\Big\},\nonumber\\
S^>(X,p)&=&\frac{1}{(2\pi)^3}\delta(p^2-m^2)\sum_{sr}\Big\{\theta(p^0)\bar{u}_s(\mathbf{p})
{u}_r(\mathbf{p})\bar{f}_q^{sr}(X,p)+\theta(-p^0)\bar{v}_s(-\mathbf{p})
{v}_r(-\mathbf{p}){f}_{\bar{q}}^{sr}(X,-p)\Big\},
\end{eqnarray}
where $\bar{f}_{{q}}^{sr}(X,p)=\delta_{sr}-{f}_{{q}}^{sr}(X,p)$ and $\bar{f}_{\bar{q}}^{sr}(X,-p)=\delta_{sr}-{f}_{\bar{q}}^{sr}(X,-p)$ can be obtained from the ensemble average of the creation and annihilation operators. $f_q^{sr}$ is an element of a $2\times2$ Hermitian matrix, which can be diagonalized to give $f_q^s$, and one can easily find that $\bar{f}_q^{sr}$ gives $1-f_q^s$ after being diagonalized. With $s=\pm1$ denoting the spin up and down along the direction set by the unit vector $n^{\pm\mu}$. The mean polarization vector (in the LAB frame) is $n^{(0)\mu}(X,p)=\theta(p^0)n^{+\mu}(X,\mathbf{p})-\theta(-p^0)n^{-\mu}(X,-\mathbf{p})$, which is a unit time-like vector satisfying $n^{(0)\mu}(X,p)n^{(0)}_{\mu}(X,p)=-1$, with 
\begin{eqnarray}
n^{\pm\mu}(X,\mathbf{p})=\pm\left(\frac{\mathbf{n}_*^\pm\cdot\mathbf{p}}{m}~,~\mathbf{n}_*^\pm+\frac{\mathbf{n}_*^\pm\cdot\mathbf{p}}{m(E_\mathbf{p}+m)}\mathbf{p}\right).
\end{eqnarray}
$\mathbf{n}_*^\pm$ is the direction of mean polarization of particles (label by +) or anti-particles (label by -) with momentum $\mathbf{p}$ measured in the Particle Rest Frame, and satisfies $\mathbf{n}_*^\pm\cdot \mathbf{n}_*^\pm=-1$. And likewise for spin decomposition of $S^>$, it can be obtained simply by taking the exchange $f_q\leftrightarrow\bar{f}_q$ and $f_{\bar{q}}\leftrightarrow\bar{f}_{\bar{q}}$. Note that $f_q^s$ is the particle distribution parallel(s=+) and antiparallel(s=-) to the unit vector $n^{\pm\mu}$. One can also introduce the vector charge distribution $f_V$ and axial charge distribution $f_A$, which are combinations of $f_q^s$, $f_q^+ + f_q^-=f_{Vq}$ and  $f_q^+ - f_q^-=f_{Aq}$. The magnitude of polarization can be defined through $f_{Aq}=f_{Vq}(X,\mathbf{p})\zeta_q(X,\mathbf{p})$. $f_q^s=f_{Vq}(1+s\zeta_q)$, the positive quantity $\zeta_{q/\bar{q}}(X,\mathbf{p})$ defines the magnitude of spin polarization. The case $\zeta_{q/\bar{q}}(X,\mathbf{p})=1$ corresponds to a pure state, while the case $\zeta_{q/\bar{q}}(X,\mathbf{p})<1$ describes a mixed state \cite{Florkowski:2019gio}. It is worth noticing that, for outgoing particles one can varify $\bar{f}_q^s=1-f_q^s$, so that $\bar{f}_V=1-f_V$ and $\bar{f}_A=-f_A$. This can also be understood from the fact that, the axial distribution function $f_A$ comes from the off-diagonal component of $f^{sr}$, and $\bar{f}^{sr}=\delta_{rs}-{f}^{sr}$ leads to $\bar{f}_A=-f_A$. With the vector charge distribution $f_V$ and axial charge distribution $f_A$, the classical components can be rewritten as 
\begin{eqnarray}
\label{Sfv}
\mathcal{V}^{(0)}_\mu(X,p)&=&\frac{2p_\mu}{(2\pi)^32E_\mathbf{p}}\Big\{\delta(p^0-E_\mathbf{p})f_{Vq}(X,\mathbf{p})+\delta(p^0+E_\mathbf{p})\bar{f}_{V\bar{q}}(X,-\mathbf{p})\Big\},\nonumber\\
\mathcal{A}^{(0)}_\mu(X,p)&=&\frac{2m}{(2\pi)^32E_\mathbf{p}}\Big\{\delta(p^0-E_\mathbf{p})n^{+}_{\mu}(X,\mathbf{p})f_{Aq}(X,\mathbf{p})-\delta(p^0+E_\mathbf{p})n^{-}_{\mu}(X,-\mathbf{p})\bar{f}_{A\bar{q}}(X,-\mathbf{p})\Big\}.
\end{eqnarray}
And likewise, for $\bar{\mathcal{V}}^{(0)}_\mu$ and $\bar{\mathcal{A}}_\mu^{(0)}$, one just takes the replacement $f_{Vq/\bar{q}}\leftrightarrow \bar{f}_{Vq/\bar{q}}$ and $f_{Aq/\bar{q}}\leftrightarrow \bar{f}_{Aq/\bar{q}}$.

Consider the transport of particle sector in $\mathcal{V}^{(0)}_\mu$ and $\mathcal{A}_\mu^{(0)}$, namely the lefthand side contains the delta function $\delta(p^0-E_{\mathbf{p}})$. Yet in $S^<(p_2)$, $S^>(p_3)$ and $S^<(p_1)$, both particle and antiparticle exist, these corresponds to different scattering process. Only three of the eight channels are allowed by the energy-momentum conservation. Each term contains a product of a combination of four quark and anti-quark distribution functions, with $f_{Vq}(X,\mathbf{p})$ at present in all the terms, being the external function under study. One may attribute a diagram to each of these processes in a loose sense, by assigning incoming quark lines to $f_{Vq}$; the incoming antiquark line to $f_{V\bar q}$; the outgoing quark lines to $\bar{f}_{Vq}$; the outgoing antiquark lines to $\bar{f}_{V\bar q}$. 
The allowed channels correspond to the quark-quark scattering as well as quark-antiquark scattering. Other channels involve particle and antiparticle creation and annihilation, and can be categorized as off-shell processes. Together with the gain term, one can obtain the transport equation of the vector charge. In the following, when considering detailed balance, we only focus on the first channel, namely the quark-quark scattering.

In kinetic theory, the local equilibrium state is specified by the distribution functions that eliminate the collision kernel. This implies that the distribution functions must depend only on the linear combination of the collisional conserved quantities: the particle number, the energy and momentum, and angular momentum.

We first focus on the collision term of $p\cdot\nabla \mathcal{S}^{(0)}$ in Eq.(\ref{TSLO}). In the following, we neglect the subscript $q$ in $f_{Vq}$, and take $f_{V}$ for the particle sector. The detailed balance requires the collision term to be vanishing, which can be further divided into terms with only vector charge distribution, term involving only axial charge distribution, as well as mixed terms with both $f_V$ and $f_A$.  Terms involving only vector charge distribution is 
\begin{eqnarray}
m(m^2+{p_2\cdot p_3})(m^2+p\cdot p_1)\big(f_{Vq}^{\mathbf{p}_1}\bar{f}_{V}^{\mathbf{p}}\bar{f}_{V}^{\mathbf{p}_2}{f}_{V}^{\mathbf{p}_3}-f_{V}^{\mathbf{p}}\bar{f}_{V}^{\mathbf{p}_1} {f}_{V}^{\mathbf{p}_2}\bar{f}_{V}^{\mathbf{p}_3}\big).
\end{eqnarray}
The vanishing of this term implies that the local equilibrium distribution $f_{Vq}^{\text{LE}}$ is the Fermi-Dirac distribution.
Requiring that the $\mathcal{A}_\mu^{(0)}$ has the structure $\mathcal{A}_\mu^{(0)}=m n_\mu f_A$, then one can easily show that terms contains only $\mathcal{A}_\mu$ vanishes in the local equilibrium, 
\begin{eqnarray}
&&m\big[(p_2\cdot n_3)(p_3\cdot n_2)(p\cdot n_1)(p_1\cdot n)
+(m^2+p_2\cdot p_3)(m^2+p\cdot p_1)(n\cdot n_1)(n_2\cdot n_3)
\\
&&-(m^2+p_2\cdot p_3)(p\cdot n_1)(p_1\cdot n)(n_3\cdot n_2)
-(m^2+p\cdot p_1)(n\cdot n_1)(p_2\cdot n_3)(p_3\cdot n_2)
\big](\bar{f}_{A}^{\mathbf{p}}f_{A}^{\mathbf{p}_1}\bar{f}_{A}^{\mathbf{p}_2}f_{A}^{\mathbf{p}_3}-{f}_{A}^{\mathbf{p}}\bar{f}_{A}^{\mathbf{p}_1}{f}_{A}^{\mathbf{p}_2}\bar{f}_{A}^{\mathbf{p}_3}).\nonumber
\end{eqnarray}
Considering that $\bar{f}_{A}=-{f}_{A}$, one has explicitly $\bar{f}_{A}^{\mathbf{p}}f_{A}^{\mathbf{p}_1}\bar{f}_{A}^{\mathbf{p}_2}f_{A}^{\mathbf{p}_3}-{f}_{A}^{\mathbf{p}}\bar{f}_{A}^{\mathbf{p}_1}{f}_{A}^{\mathbf{p}_2}\bar{f}_{A}^{\mathbf{p}_3}=0$. This does not have any restriction on the equilibrium distribution function. Finally, terms involving mixture of $f_{V}$ and $f_A$ is 
\begin{eqnarray}
&&m(m^2+p_2\cdot p_3)((p\cdot n^{1}) (p_{1}\cdot n)-(m^2+{p\cdot p_1}) n_1\cdot n)(\bar{f}_{V}^{\mathbf{p}_2}{f}_{V}^{\mathbf{p}_3} -{f}_{V}^{\mathbf{p}_2}\bar{f}_{V}^{\mathbf{p}_3}) {f}_{A}^{\mathbf{p}} {f}_{A}^{\mathbf{p}_1}\nonumber\\
&+&m(m^2+p\cdot p_1)((p_2\cdot n^{3})(p_3\cdot n^2)-(m^2+{p_2\cdot p_3})n_2\cdot n_3)(\bar{f}_{V}^{\mathbf{p}_1} {f}_{V}^{\mathbf{p}}-{f}_{V}^{\mathbf{p}_1}\bar{f}_{V}^{\mathbf{p}}){f}_{A}^{\mathbf{p}_2}{f}_{A}^{\mathbf{p}_3}.
\end{eqnarray}
A trivial solution is $f_{Aq}^{\text{LE}}=0$, indicating that at classical level, the axial charge distribution function has only trivial solution, if the system is not initially polarized. 

We then consider the collision term in the transport equation $p\cdot\nabla\mathcal{A}^{(0)}_\mu$ in Eq.(\ref{TSLO}). Terms in the momentum integral can be simplified to 
\begin{eqnarray}
&-&m(m^2+p_2\cdot p_3)(m^2+p_1\cdot p) n_\mu{f}_{A}^{\mathbf{p}}(f_V^{\mathbf{p}_1} \bar{f}_{V}^{\mathbf{p}_2}{f}_{V}^{\mathbf{p}_3}+\bar{f}_{V}^{\mathbf{p}_1} {f}_{V}^{\mathbf{p}_2}\bar{f}_{V}^{\mathbf{p}_3})\nonumber\\
&+&m(m^2+p_2\cdot p_3)\left[(m^2+p_1\cdot p)n_\mu^1-(p\cdot n^1)(p_\mu+p_{1\mu})\right]f_{A}^{\mathbf{p}_1}(\bar{f}_{V}^{\mathbf{p}}\bar{f}_{V}^{\mathbf{p}_2}{f}_{V}^{\mathbf{p}_3}+{f}_{V}^{\mathbf{p}}{f}_{V}^{\mathbf{p}_2}\bar{f}_{V}^{\mathbf{p}_3})\\
&+&m(m^2+p_1\cdot p)n_\mu\left[(p_2\cdot n^{3})(p_3\cdot n^2)-(m^2+{p_2\cdot p_3}) n_2\cdot n_3\right]{f}_{A}^{\mathbf{p}} {f}_{A}^{\mathbf{p}_2}{f}_{A}^{\mathbf{p}_3}\nonumber\\
&-&m\big[(m^2+p_1\cdot p)n^1_\mu-(p_\mu+p_{1\mu})p\cdot n^1\big]\big[(p_2\cdot n^{3})(p_3\cdot n^2)-(m^2+p_2\cdot p_3)n^2\cdot n^{3}\big]f_{A}^{\mathbf{p}_1}{f}_{A}^{\mathbf{p}_2}f_{A}^{\mathbf{p}_3},\nonumber
\end{eqnarray}
where we have considered that $\bar{f}_V+f_V=1$ and $\bar{f}_A=-f_A$. The vanishing of the collision terms again requires that the classical level of axial charge distribution has trivial solution $f_{Aq}^{\text{LE}}=0$. Considering the relations between spin components of the Wigner function in Eq.(\ref{Spincomponents}), the above solution leads to a nonzero classical scalar component $\mathcal{S}^{(0)}_{\text{LE}}\neq0$ and a vanishing tensor component $\mathcal{S}_{\text{LE},\mu\nu}^{(0)}=0$. The pseudo-scalar at $\mathcal{O}(\hbar)$ is directly shown to be vanished by the constraint equations, $\mathcal{P}^{(0)}=0$. Since the axial charge current appears at $\mathcal{O}(\hbar)$ order, it is necessary to analyze the first order transport equation. In an initially unpolarized system, when considering the fermionic 2 by 2 collisions, the spin polarization can be produced as a quantum effect. The equilibrium spin polarization is then derived from the detailed balance of the collision term in the first order transport equation. 

\subsection{Collision Term at Quantum Level}
In a system without background electromagnetic field and vorticity field, one has $\mathcal{A}^{(0)}_{\text{LE},\mu}=0$, this greatly simplifies the transport equations Eq.(\ref{TV1}) and Eq.(\ref{TA1}). Besides, one can easily varify that, within the NJL model, the vanishing $\mathcal{A}^{(0)}$ leads to vanishing $\Sigma_A^{(0)}$ and $\Sigma_P^{(0)}$. In the leading order in the $1/N_c$ expansion, $\bar{\Sigma}_T^{(0)\alpha\beta}$ can also be shown to vanish. With the conditions above, the vector component and axial-vector component are still on-shell at $\mathcal{O}(\hbar)$ order, that $(p^2-m^2)\mathcal{V}_\mu^{(1)}=0$ and $(p^2-m^2)\mathcal{A}_\mu^{(1)}=0$. The orthogonal relations are unchanged as well that $p_{[\mu}\mathcal{V}_{\nu]}^{(1)}=0$ and $p^\mu\mathcal{A}_\mu^{(1)}=0$. The transport equations of $\mathcal{V}^{(1)}_\mu$ to the first order of $\hbar$ (\ref{TV1}) can be simplified to give 
\begin{eqnarray}
\label{TV1LE}
(p\cdot\nabla)\mathcal{V}^{(1)}_\mu
&=&I_{V,\text{gain}}^{(1)}-I_{V,\text{loss}}^{(1)}
=
m\big(\widehat{\Sigma_S^{(0)}\mathcal{V}_{\mu}^{(1)}}+\widehat{\Sigma_S^{(1)}\mathcal{V}_{\mu}^{(0)}}\big)
+p^\nu\big(\widehat{\Sigma_{V\nu}^{(0)}\mathcal{V}_{\mu}^{(1)}}
+\widehat{\Sigma_{V\nu}^{(1)}\mathcal{V}_{\mu}^{(0)}}\big).
\end{eqnarray}
It is obvious that, at $\mathcal{O}(\hbar)$ the transport equation of $\mathcal{V}^{(1)}_\mu$ still contains only diffusion terms. The transport equations of $\mathcal{A}^{(1)}_\mu$ to the first order of $\hbar$ can be simplified to give 
\begin{eqnarray}
\label{TA1LE}
(p\cdot\nabla)\mathcal{A}^{(1)}_\mu
&=&I_{A,\text{gain}}^{(1)}-I_{A,\text{loss}}^{(1)}\\
&=&m\widehat{{\Sigma}_S^{(0)}\mathcal{A}^{(1)}_{\mu}}+p_\nu\widehat{{\Sigma}_{V}^{(0)\nu}\mathcal{A}^{(1)}_{\mu}}
-\frac{1}{2}\epsilon_{\mu\nu\rho\sigma}(\nabla^\sigma \widehat{{\Sigma}_V^{(0)\nu})\mathcal{V}^{(0)\rho}}+p_\nu\widehat{{\Sigma}_{A\mu}^{(1)}\mathcal{V}^{(0)\nu}}-p_\mu\widehat{{\Sigma}_{A\nu}^{(1)}\mathcal{V}^{(0)\nu}}
+\frac{m}{2}\epsilon_{\rho\nu\lambda\mu}\widehat{{\Sigma}_T^{(1)\rho\nu}\mathcal{V}^{(0)\lambda}}.\nonumber
\end{eqnarray}
The first two terms $m\widehat{{\Sigma}_S^{(0)}\mathcal{A}^{(1)}_{\mu}}+p_\nu\widehat{{\Sigma}_{V}^{(0)\nu}\mathcal{A}^{(1)}_{\mu}}$ can be understood as spin diffusion effect, terms with same structure but of $\mathcal{O}(\hbar^{0})$ also appears in the transport equation of $\mathcal{A}_\mu^{(0)}$. The third term $-\frac{1}{2}\epsilon_{\mu\nu\rho\sigma}(\nabla^\sigma \widehat{{\Sigma}_V^{(0)\nu})\mathcal{V}^{(0)\rho}}$ contains only the vector components of the Wigner function, indicating that in an initially unpolarized system, non-zero spin polarization $\mathcal{A}^{(1)}\neq0$ can be generated through interacting with the matter. Consider the leading order in $1/N_c$ expansion, the self-energy $\Sigma^>_{\text{LO}}$ is defined by
\begin{eqnarray}
{\Sigma}^{>(0)}(X,p)
=&&G^2\int_{qk}~{S}^{>(0)}(X,p+q)\text{Tr}\big[S^{<(0)}(X,k+q){S}^{>(0)}(X,k)\big],\nonumber\\
{\Sigma}^{>(1)}(X,p)
=&&G^2\int_{qk}~{S}^{>(1)}(X,p+q)\text{Tr}\big[S^{<(0)}(X,k+q){S}^{>(0)}(X,k)\big]\nonumber\\
&+&G^2\int_{qk}~{S}^{>(0)}(X,p+q)\text{Tr}\big[S^{<(1)}(X,k+q){S}^{>(0)}(X,k)\big]\nonumber\\
&+&G^2\int_{qk}~{S}^{>(0)}(X,p+q)\text{Tr}\big[S^{<(0)}(X,k+q){S}^{>(1)}(X,k)\big].
\end{eqnarray}
In the transport equation, the components $\Sigma_S^{(0)}$,  $\Sigma_V^{(0)}$, $\Sigma_S^{(1)}$,  $\Sigma_V^{(1)}$ and $\Sigma_A^{(1)}$, $\Sigma_T^{(1)}$ are involved. Under the expansion to $1/N_c$ order, taking $\mathcal{S}$ and $\mathcal{A}_\mu$ as independent components, and considering the relations between the spin components (\ref{Spincomponents}), the self-energy components can be evaluated. We list only $\Sigma_T^{(1)}$ in the following, other components are easier to calculate, 
\begin{eqnarray}
\label{selfEcompo2}
\bar{\Sigma}^{(1)}_{T\mu\nu}(p)
&=&G^2\int_{qk}~\Big(1+\frac{(k+q)\cdot k}{m^2}\Big)\frac{(p+q)_{\nu}}{2m^2}\mathcal{S}^{(0)}(k+q)\bar{\mathcal{S}}^{(0)}(k)\nabla_{\mu}\bar{\mathcal{S}}^{(0)}(p+q)\nonumber\\
&-&G^2\int_{qk}~\Big(1+\frac{(k+q)\cdot k}{m^2}\Big)\frac{(p+q)_{\mu}}{2m^2}\mathcal{S}^{(0)}(k+q)\bar{\mathcal{S}}^{(0)}(k)\nabla_{\nu}\bar{\mathcal{S}}^{(0)}(p+q)\nonumber\\
&-&G^2\int_{qk}~\Big(1+\frac{(k+q)\cdot k}{m^2}\Big)\frac{1}{m}\mathcal{S}^{(0)}(k+q)\bar{\mathcal{S}}^{(0)}(k)\epsilon_{\rho\sigma\mu\nu}(p+q)^\rho\bar{\mathcal{A}}^{(1)\sigma}(p+q).
\end{eqnarray}
With explicit components of the self-energies, the loss term on the righthand side of Eq.(\ref{TV1LE}) is given by
\begin{eqnarray}
I_{V,\text{loss}}^{(1)}&=&G^2p^\mu\int_{qk}~\Big(1+\frac{(k+q)\cdot k}{m^2}\Big)\Big(m+\frac{p\cdot (p+q)}{m}\Big)\nonumber\\
&&\times\Big\{\mathcal{S}^{(0)}(k+q)\bar{\mathcal{S}}^{(0)}(k)\bar{\mathcal{S}}^{(0)}(p+q){\mathcal{S}}^{(1)}(p)+\mathcal{S}^{(1)}(k+q)\bar{\mathcal{S}}^{(0)}(k)\bar{\mathcal{S}}^{(0)}(p+q){\mathcal{S}}^{(0)}(p)\nonumber\\
&&~+\mathcal{S}^{(0)}(k+q)\bar{\mathcal{S}}^{(1)}(k)\bar{\mathcal{S}}^{(0)}(p+q){\mathcal{S}}^{(0)}(p)+\mathcal{S}^{(0)}(k+q)\bar{\mathcal{S}}^{(0)}(k)\bar{\mathcal{S}}^{(1)}(p+q){\mathcal{S}}^{(0)}(p)\Big\},
\end{eqnarray}
with $f_V^{\text{LE}(0)}$ taking the Fermi-Dirac distribution function at local equilibrium, $\mathcal{S}^{(1)}(p)$ adopts the simple vanishing solution that $\mathcal{S}^{(1)}(p)=0$ at local equilibrium.  

Likewise, the loss term on the righthand side of Eq.(\ref{TA1LE}) is, 
\begin{eqnarray}
\label{IAloss}
I_{A,\text{loss}}^{(1)}&=&G^2\int_{qk}\Big\{\Big(1+\frac{(k+q)\cdot k}{m^2}\Big)\Big(m+\frac{p\cdot (p+q)}{m}\Big)\mathcal{S}^{(0)}(k+q)\bar{\mathcal{S}}^{(0)}(k)\bar{\mathcal{S}}^{(0)}(p+q)\mathcal{A}^{(1)}_{\mu}(p)\nonumber\\
&&\qquad\quad+\Big(1+\frac{(k+q)\cdot k}{m^2}\Big)\Big(m+\frac{p\cdot (p+q)}{m}\Big)\mathcal{S}^{(0)}(k+q)\bar{\mathcal{S}}^{(0)}(k)\bar{\mathcal{A}}_\mu^{(1)}(p+q)\mathcal{S}^{(0)}(p)\nonumber\\
&&\qquad\quad-\Big(1+\frac{(k+q)\cdot k}{m^2}\Big)\frac{(2p+q)_\mu p^\nu}{m} \mathcal{S}^{(0)}(k+q)\bar{\mathcal{S}}^{(0)}(k)\mathcal{S}^{(0)}(p)\bar{\mathcal{A}}^{(1)}_{\nu}(p+q)\nonumber\\
&&\qquad\quad+\Big(1+\frac{(k+q)\cdot k}{m^2}\Big)\epsilon_{\mu\nu\rho\sigma}\frac{(p+q)^\nu p^\rho}{2m^2}\mathcal{S}^{(0)}(k+q)\bar{\mathcal{S}}^{(0)}(k)\bar{\mathcal{S}}^{(0)}(p+q)\big[\nabla^\sigma\mathcal{S}^{(0)}(p)\big]\nonumber\\
&&\qquad\quad-\Big(1+\frac{(k+q)\cdot k}{m^2}\Big)\epsilon_{\mu\nu\rho\sigma}\frac{(p+q)^{\nu}p^\rho}{2m^2}\mathcal{S}^{(0)}(k+q)\bar{\mathcal{S}}^{(0)}(k)\mathcal{S}^{(0)}(p)\big[\nabla^{\sigma}\bar{\mathcal{S}}^{(0)}(p+q)\big]
\Big\}.
\end{eqnarray}
The first order component $\mathcal{A}^{(1)}_\mu$ appears in the first three lines, the last two lines contains only the classical scalar component $\mathcal{S}^{(0)}$ and its derivative. The appearance of terms involving purely classical scalar component $\mathcal{S}^{(0)}$ indicates that spin polarization $\mathcal{A}^{(1)}_\mu$ can be generated by collisions. Since such terms also involves spatial derivative of $\mathcal{S}^{(0)}$, the spin polarization does not appear in an homogeneous system. When the system achieves local equilibrium, the detailed balance requires that the gain term and loss term cancels out. In the following, we show that terms containing purely scalar components would contribute to the equilibrium expression of $\mathcal{A}^{(1)}_\mu$, and is related to local vorticity. With the classical expression of $\mathcal{S}^{(0)}$ in Eq.(\ref{Sfv}), and considering only the particle part in $\mathcal{S}^{(0)}$ and $\mathcal{A}^{(1)}_\mu$, namely terms with $\delta(p^0-E_p)$. Using the Schouten identity $p^\lambda\epsilon_{\mu\nu\rho\sigma}+p^\mu\epsilon_{\nu\rho\sigma\lambda}+p^\nu\epsilon_{\rho\sigma\lambda\mu}+p^\rho\epsilon_{\sigma\lambda\mu\nu}+p^\sigma\epsilon_{\lambda\mu\nu\rho}=0$
to simplify terms with spatial derivatives in (\ref{IAloss}), for instance, 
\begin{eqnarray}
\epsilon_{\mu\nu\rho\sigma}(p+q)^\nu p^\rho\big[\nabla^\sigma f_V(X,p)\big]
&=&
-p_\mu\epsilon_{\nu\rho\sigma\lambda}(p+q)^\nu p^\rho \nabla^\sigma \beta^\lambda f'_V(X,p)
-m^2\epsilon_{\mu\nu\sigma\lambda}(2p+q)^\nu \nabla^\sigma \beta^\lambda f'_V(X,p)
\nonumber\\
&&
-\left(p\cdot(p+q)+m^2\right) \epsilon_{\nu\sigma\lambda\mu}p^\nu \nabla^\sigma \beta^\lambda f'_V(X,p),
\end{eqnarray}
where $f'(p)=\frac{d}{d(\beta\cdot p)}f$. And similarly for the other derivative term $-\epsilon_{\mu\nu\rho\sigma}(p+q)^{\nu}p^\rho[\nabla^{\sigma}\bar{f}_V^{(0)}(p+q)]$. Considering the fact that $\bar{\mathcal{A}}^{(1)}_\mu=-{\mathcal{A}}^{(1)}_\mu$ and that $\bar{f}_V=1-f_V$, the collision term could be rearranged into two parts, and simplified as
\begin{eqnarray}
\label{collisionterm}
I_{A,\text{gain}}^{(1)}-I_{A,\text{loss}}^{(1)}&=&G^2\int_{qk}\{F_1+F_2\},
\end{eqnarray}
with $F_1$ stands for 
\begin{eqnarray}
F_1=&+&\frac{[m^2+{(k+q)\cdot k}][m^2+p\cdot (p+q)]}{[(2\pi)^3]^3E_{k+p}E_{k}E_{p+q}}\Big\{f_V^{k+q}\bar{f}_V^{k}\bar{f}_V^{p+q}+\bar{f}_V^{k+q}{f}_V^{k}{f}_V^{p+q}\Big\}\nonumber\\
&&\times\Big[\mathcal{A}^{(1)}_{\mu}(p)+\frac{1}{(2\pi)^32E_p}\epsilon_{\mu\nu\sigma\lambda}p^\nu \nabla^\sigma \beta^\lambda f'_V(p)\Big]\nonumber\\
&-&\frac{[m^2+{(k+q)\cdot k}][m^2+p\cdot (p+q)]}{[(2\pi)^3]^3E_{k+p}E_{k}E_{p}}\Big\{f_V^{k+q}\bar{f}_V^{k}f_V^{p}+\bar{f}_V^{k+q}{f}_V^{k}\bar{f}_V^{p}\Big\}\nonumber\\
&&\times\Big[{\mathcal{A}}_\mu^{(1)}(p+q)+\frac{1}{(2\pi)^32E_{p+q}}\epsilon_{\mu\nu\sigma\lambda}(p+q)^{\nu}\nabla^{\sigma}\beta^\lambda {f}'_V(p+q)\Big], 
\end{eqnarray}
where $f_V^p$ is the shorthand notation for $f_V^{(0)}(X,p)$. Notice that the solution $\mathcal{A}^{(1)}_{\mu}(p)=-\frac{1}{(2\pi)^32E_p}\epsilon_{\mu\nu\sigma\lambda}p^\nu \nabla^\sigma \beta^\lambda f'_V(p)$ would lead to vanishing $F_1$. In the following, we show that this solution of $\mathcal{A}^{(1)}_\mu$ also makes $F_2$ vanishing, so the collision term vanishes as required by the detailed balance. $F_2$ in Eq.(\ref{collisionterm}) is given by 
\begin{eqnarray}
F_2&=&-\frac{[m^2+{(k+q)\cdot k}]}{[(2\pi)^3]^4E_{k+p}E_{k}E_{p+q}E_p}\frac{1}{2}p_\mu\epsilon_{\nu\rho\sigma\lambda}(p+q)^\nu p^\rho \nabla^\sigma \beta^\lambda \Big\{f_V^{k+q}\bar{f}_V^k\bar{f}_V^{p+q}+ \bar{f}_V^{k+q} {f}_V^k {f}_V^{p+q}\Big\}f'_V(X,p)\nonumber\\
&&-\frac{m^2[m^2+(k+q)\cdot k]}{[(2\pi)^3]^4E_{k+p}E_{k}E_{p+q}E_p}\frac{1}{2}\epsilon_{\mu\nu\sigma\lambda}(2p+q)^\nu \nabla^\sigma \beta^\lambda\Big\{f_V^{k+q}\bar{f}_V^k\bar{f}_V^{p+q}+\bar{f}_V^{k+q}{f}_V^{k}{f}_V^{p+q}\Big\}f'_V(X,p)\nonumber\\
&&-\frac{[m^2+{(k+q)\cdot k}]}{[(2\pi)^3]^4E_{k+p}E_{k}E_{p+q}E_p}\frac{1}{2}(p+q)_\mu\epsilon_{\nu\rho\sigma\lambda}(p+q)^{\nu}p^\rho\nabla^{\sigma}\beta^\lambda \Big\{f_V^{k+q}\bar{f}_V^{k}f_V^{p}+\bar{f}_V^{k+q}{f}_V^{k}\bar{f}_V^{p}\Big\}{f}'_V(X,p+q)\nonumber\\
&&+\frac{m^2[m^2+{(k+q)\cdot k}]}{[(2\pi)^3]^4E_{k+p}E_{k}E_{p+q}E_p}\frac{1}{2}\epsilon_{\mu\nu\sigma\lambda}(2p+q)^{\nu}\nabla^{\sigma}\beta^\lambda \Big\{f_V^{k+q}\bar{f}_V^{k}f_V^{p}+\bar{f}_V^{k+q}{f}_V^{k}\bar{f}_V^{p}\Big\}{f}'_V(X,p+q)\nonumber\\
&&+\frac{[m^2+(k+q)\cdot k]}{[(2\pi)^3]^3E_{k+p}E_{k}E_p}(2p+q)_\mu p^\nu\Big\{f_V^{k+q}\bar{f}_V^{k}f_V^{p}+\bar{f}_V^{k+q}{f}_V^{k}\bar{f}_V^{p}\Big\}{\mathcal{A}}^{(1)}_{\nu}(p+q).
\end{eqnarray}
Substituting the solution of $\mathcal{A}_\mu^{(1)}$, $F_2$ can be simplified to 
\begin{eqnarray}
F_2&=&-\frac{[2m^2+q\cdot k]}{[(2\pi)^3]^4E_{k+q}E_{k}E_{p+q}E_p}\left\{\frac{m^2}{2}\epsilon_{\mu\nu\sigma\lambda}(2p+q)^{\nu}\nabla^{\sigma}\beta^\lambda +\frac{p_\mu}{2}\epsilon_{\nu\rho\sigma\lambda}q^{\nu}p^\rho\nabla^{\sigma}\beta^\lambda \right\}\nonumber\\
&&\times\frac{d}{d(\beta\cdot p)}\Big\{f_V^{k+q}\bar{f}_V^{k}\bar{f}_V^{p+q}f_V^{p}-\bar{f}_V^{k+q}{f}_V^{k}{f}_V^{p+q}\bar{f}_V^{p}\Big\}=0.
\end{eqnarray}
The last term vanishes because of the detailed balance of number distribution function. Thus the detailed balance gives the local-equilibrium distribution of spin distribution function 
\begin{eqnarray}
\mathcal{A}_{\mu}^{\text{LE}}(p)=\mathcal{A}^{\text{LE}(0)}_{\mu}(p)+\hbar\mathcal{A}^{\text{LE}(1)}_{\mu}(p)&=&-\frac{\hbar}{(2\pi)^32E_p}\epsilon_{\mu\nu\sigma\lambda}p^\nu \nabla^\sigma \beta^\lambda f'_{V,\text{LE}}(X,p).
\end{eqnarray}
This solution indicates that in an initially unpolarized system, non-zero spin polarization can be generated from the coupling between vector and axial-vector charges. The equilibrium spin polarization is found to be created by a thermal vorticity and is orthogonal to the momentum. This equilibrium solution is self-consistently obtained from the detailed balance and agrees with the results in previous research \cite{Becattini:2013fla, Fang:2016vpj, Gao:2018jsi}.

\section{Conclusion}
\label{s4}
Spin is a quantum effect and is normally neglected in a classical transport theory. In this work, we addressed the problem of spin polarization in the Wigner function formalism of quantum kinetic theory. While non-equilibrium distributions are related to the details of the interaction of the system, namely the collision terms, the corresponding local equilibrium distributions are determined only by the detailed balance between the loss and gain terms, namely the disappearing of the collision terms. We obtained the local equilibrium spin distribution by requiring the detailed balance for the Kadanoff-Baym equations. To be specific, we take the Nambu--Jona-Lasinio interaction as an example to calculate the collision terms in the constraint and transport equations at classical level and to the first order in $\hbar$. We found that, for an initially non-polarized system without external electromagnetic fields, while the local equilibrium spin distribution is trivial at classical level, the quantum correction internally generated by the inhomogeneous vorticity of the system leads to a non-trivial spin distribution. Our result supports the statement in previous studies.   
\\

\noindent  {\bf Acknowledgement}: The authors would like to thank Drs. Shi Pu, Shuzhe Shi and Xinli Seng for fruitful discussion. The work is supported by the NSFC Grant Nos. 11890712 and 11905066. ZyW is also supported by the Postdoctoral Innovative Talent Support Program of Tsinghua University. 

\begin{appendix}
\section{Spin decomposition \& Semiclassical expansion}
\label{a1}
The collision terms in the Kadanoff-Baym equations (\ref{KBequation}), $[\Sigma^<,S^>]_\star-[\Sigma^>,S^<]_\star$ and $\{\Sigma^<,S^>\}_\star-\{\Sigma^>,S^<\}_\star$, are $4\times4$ matrices which should be docomposed with the Clifford algebra. To the lowest order of $\hbar$ the loss terms can be decomposed as 
\begin{eqnarray}
[\Sigma^>,S^<]^{(0)}_\star&=&
+2i\Big(\bar{\Sigma}_{V\mu}\mathcal{A}^\mu
-\bar{\Sigma}_{A\mu}\mathcal{V}^\mu\Big)i\gamma^5\nonumber\\
&&
+2i\Big(\bar{\Sigma}_P\mathcal{A}_\mu
+\bar{\Sigma}_{V}^\nu\mathcal{S}_{\nu\mu}
-\bar{\Sigma}_{A\mu}\mathcal{P}
+\bar{\Sigma}_{T\mu\nu}\mathcal{V}^\nu\Big)\gamma^\mu\nonumber\\
&&
+2i\Big(\bar{\Sigma}_P\mathcal{V}_\mu-\bar{\Sigma}_{V\mu}\mathcal{P}+\bar{\Sigma}_{A}^\nu\mathcal{S}_{\nu\mu}+\bar{\Sigma}_{T\mu\nu}\mathcal{A}^\nu\Big)\gamma^5\gamma^\mu\nonumber\\
&&
+2i\Big(\bar{\Sigma}_{A[\mu}\mathcal{A}_{\nu]}-\bar{\Sigma}_{V[\mu}\mathcal{V}_{\nu]}-\bar{\Sigma}_{T\alpha[\mu}\mathcal{S}_{~\nu]}^{\alpha}\Big)\frac{\sigma^{\mu\nu}}{2},
\end{eqnarray}
and 
\begin{eqnarray}
\{\Sigma^>,S^<\}^{(0)}_\star
&=&
+2\Big(\bar{\Sigma}_S\mathcal{S}-\bar{\Sigma}_P\mathcal{P}+\bar{\Sigma}_{V\mu}\mathcal{V}^\mu-\bar{\Sigma}_{A\mu}\mathcal{A}^\mu+\frac{1}{2}\bar{\Sigma}_{T\mu\nu}\mathcal{S}^{\mu\nu}\Big)\nonumber\\
&&
+2\Big(\bar{\Sigma}_S\mathcal{P}+\bar{\Sigma}_P\mathcal{S}+\frac{1}{4}\epsilon^{\mu\nu\alpha\beta}\bar{\Sigma}_{T\mu\nu}\mathcal{S}_{\alpha\beta}\Big)i\gamma^5\nonumber\\
&&
+2\Big(\bar{\Sigma}_S\mathcal{V}_\mu
+\bar{\Sigma}_{V\mu}\mathcal{S}
+\frac{1}{2}\epsilon_{\sigma\nu\lambda\mu}(\bar{\Sigma}_{A}^\sigma\mathcal{S}^{\nu\lambda}+\bar{\Sigma}_{T}^{\sigma\nu}\mathcal{A}^\lambda)\Big)\gamma^\mu\nonumber\\
&&
+2\Big(\bar{\Sigma}_S\mathcal{A}_\mu
+\bar{\Sigma}_{A\mu}\mathcal{S}
+\frac{1}{2}\epsilon_{\sigma\nu\lambda\mu}(\bar{\Sigma}_{V}^\sigma\mathcal{S}^{\nu\lambda}
+\bar{\Sigma}_T^{\sigma\nu}\mathcal{V}^\lambda)\Big)\gamma^5\gamma^\mu\nonumber\\
&&
+2\Big(\bar{\Sigma}_S\mathcal{S}_{\mu\nu}
+\bar{\Sigma}_{T\mu\nu}\mathcal{S}
+\epsilon_{\mu\nu\alpha\beta}\big(\bar{\Sigma}_{A}^\alpha\mathcal{V}^\beta
-\bar{\Sigma}_{V}^\alpha\mathcal{A}^\beta
-\frac{1}{2}\bar{\Sigma}_{T}^{\alpha\beta}\mathcal{P}
-\frac{1}{2}\bar{\Sigma}_P\mathcal{S}^{\alpha\beta}\big)\Big)\frac{\sigma^{\mu\nu}}{2}.
\end{eqnarray}
To the first order of $\hbar$ there are
\begin{eqnarray}
[\Sigma^>,S^<]^{(1)}_\star&=&
+i2\hbar\Big(\bar{\Sigma}_{V\mu}\mathcal{A}^{\mu}
-\bar{\Sigma}_{A\mu}\mathcal{V}^{\mu}\Big)^{(1)}i\gamma^5\nonumber\\
&&
+i2\hbar\Big(\bar{\Sigma}_P\mathcal{A}_\mu
+\bar{\Sigma}_{V}^\nu\mathcal{S}_{\nu\mu}
-\bar{\Sigma}_{A\mu}\mathcal{P}
+\bar{\Sigma}_{T\mu\nu}\mathcal{V}^{\nu}\Big)^{(1)}\gamma^\mu\nonumber\\
&&
+i2\hbar\Big(\bar{\Sigma}_P\mathcal{V}_\mu
-\bar{\Sigma}_{V\mu}\mathcal{P}
+\bar{\Sigma}_{A}^\nu\mathcal{S}_{\nu\mu}
+\bar{\Sigma}_{T\mu\nu}\mathcal{A}^{\nu}\Big)^{(1)}\gamma^5\gamma^\mu\nonumber\\
&&
+i2\hbar\Big(\bar{\Sigma}_{A[\mu}\mathcal{A}_{\nu]}-\bar{\Sigma}_{V[\mu}\mathcal{V}_{\nu]}-\bar{\Sigma}_{T\alpha[\mu}\mathcal{S}_{~\nu]}^{\alpha}\Big)^{(1)}\frac{\sigma^{\mu\nu}}{2}\nonumber\\
&&+i\hbar\Big(\bar{\Sigma}_S\mathcal{S}-\bar{\Sigma}_P\mathcal{P}+\bar{\Sigma}_{V\mu}\mathcal{V}^\mu-\bar{\Sigma}_{A\mu}\mathcal{A}^\mu+\frac{1}{2}\bar{\Sigma}_{T\mu\nu}\mathcal{S}^{\mu\nu}\Big)_{\text{P.B.}}\nonumber\\
&&+i\hbar\Big(\bar{\Sigma}_S\mathcal{P}+\bar{\Sigma}_P\mathcal{S}+\frac{1}{4}\epsilon^{\mu\nu\alpha\beta}\bar{\Sigma}_{T\mu\nu}\mathcal{S}_{\alpha\beta}\Big)_{\text{P.B.}}i\gamma^5\nonumber\\
&&
+i\hbar\Big(\bar{\Sigma}_S\mathcal{V}_\mu
+\bar{\Sigma}_{V\mu}\mathcal{S}
+\frac{1}{2}\epsilon_{\sigma\nu\lambda\mu}(\bar{\Sigma}_{A}^\sigma\mathcal{S}^{\nu\lambda}+\bar{\Sigma}_{T}^{\sigma\nu}\mathcal{A}^\lambda)\Big)_{\text{P.B.}}\gamma^\mu\nonumber\\
&&+i\hbar\Big(\bar{\Sigma}_S\mathcal{A}_\mu
+\bar{\Sigma}_{A\mu}\mathcal{S}
+\frac{1}{2}\epsilon_{\sigma\nu\lambda\mu}(\bar{\Sigma}_{V}^\sigma\mathcal{S}^{\nu\lambda}
+\bar{\Sigma}_T^{\sigma\nu}\mathcal{V}^\lambda)\Big)_{\text{P.B.}}\gamma^5\gamma^\mu\nonumber\\
&&
+i\hbar\Big(\bar{\Sigma}_S\mathcal{S}_{\mu\nu}
+\bar{\Sigma}_{T\mu\nu}\mathcal{S}
+\epsilon_{\mu\nu\alpha\beta}\big(\bar{\Sigma}_{A}^\alpha\mathcal{V}^\beta
-\bar{\Sigma}_{V}^\alpha\mathcal{A}^\beta
-\frac{1}{2}\bar{\Sigma}_{T}^{\alpha\beta}\mathcal{P}
-\frac{1}{2}\bar{\Sigma}_P\mathcal{S}^{\alpha\beta}\big)\Big)_{\text{P.B.}}\frac{\sigma^{\mu\nu}}{2},
\end{eqnarray}
where $(AB)^{(1)}=A^{(1)}B^{(0)}+A^{(0)}B^{(1)}$ stands for the $\mathcal{O}(\hbar)$ component of $(AB)$, and 
\begin{eqnarray}
\{\Sigma^>,S^<\}^{(1)}_\star&=&-\hbar\Big(\bar{\Sigma}_{V\mu}\mathcal{A}^\mu
-\bar{\Sigma}_{A\mu}\mathcal{V}^\mu\Big)_{\text{P.B.}}i\gamma^5\nonumber\\
&&
-\hbar\Big(\bar{\Sigma}_P\mathcal{A}_\mu
+\bar{\Sigma}_{V}^\nu\mathcal{S}_{\nu\mu}
-\bar{\Sigma}_{A\mu}\mathcal{P}
+\bar{\Sigma}_{T\mu\nu}\mathcal{V}^\nu\Big)_{\text{P.B.}}\gamma^\mu\nonumber\\
&&-\hbar\Big(\bar{\Sigma}_P\mathcal{V}_\mu-\bar{\Sigma}_{V\mu}\mathcal{P}+\bar{\Sigma}_{A}^\nu\mathcal{S}_{\nu\mu}+\bar{\Sigma}_{T\mu\nu}\mathcal{A}^\nu\Big)_{\text{P.B.}}\gamma^5\gamma^\mu\nonumber\\
&&-\hbar\Big(\bar{\Sigma}_{A[\mu}\mathcal{A}_{\nu]}-\bar{\Sigma}_{V[\mu}\mathcal{V}_{\nu]}-\bar{\Sigma}_{T\alpha[\mu}\mathcal{S}_{~\nu]}^{\alpha}\Big)_{\text{P.B.}}\frac{\sigma^{\mu\nu}}{2}\nonumber\\
&&+2\hbar\Big(\bar{\Sigma}_S\mathcal{S}-\bar{\Sigma}_P\mathcal{P}+\bar{\Sigma}_{V\mu}\mathcal{V}^{\mu}-\bar{\Sigma}_{A\mu}\mathcal{A}^{\mu}+\frac{1}{2}\bar{\Sigma}_{T\mu\nu}\mathcal{S}^{\mu\nu}\Big)^{(1)}\nonumber\\
&&
+2\hbar\Big(\bar{\Sigma}_S\mathcal{P}+\bar{\Sigma}_P\mathcal{S}+\frac{1}{4}\epsilon^{\mu\nu\alpha\beta}\bar{\Sigma}_{T\mu\nu}\mathcal{S}_{\alpha\beta}\Big)^{(1)}i\gamma^5\nonumber\\
&&
+2\hbar\Big(\bar{\Sigma}_S\mathcal{V}_\mu
+\bar{\Sigma}_{V\mu}\mathcal{S}
+\frac{1}{2}\epsilon_{\sigma\nu\lambda\mu}(\bar{\Sigma}_{A}^\sigma\mathcal{S}^{\nu\lambda}+\bar{\Sigma}_{T}^{\sigma\nu}\mathcal{A}^{\lambda})\Big)^{(1)}\gamma^\mu\nonumber\\
&&
+2\hbar\Big(\bar{\Sigma}_S\mathcal{A}_\mu
+\bar{\Sigma}_{A\mu}\mathcal{S}
+\frac{1}{2}\epsilon_{\sigma\nu\lambda\mu}(\bar{\Sigma}_{V}^\mu\mathcal{S}^{\nu\lambda}
+\bar{\Sigma}_T^{\sigma\nu}\mathcal{V}^{\lambda})\Big)^{(1)}\gamma^5\gamma^\mu\nonumber\\
&&
+2\hbar\Big(\bar{\Sigma}_S\mathcal{S}_{\mu\nu}
+\bar{\Sigma}_{T\mu\nu}\mathcal{S}
+\epsilon_{\mu\nu\alpha\beta}\big(\bar{\Sigma}_{A}^\alpha\mathcal{V}^\beta
-\bar{\Sigma}_{V}^\alpha\mathcal{A}^\beta
-\frac{1}{2}\bar{\Sigma}_{T}^{\alpha\beta}\mathcal{P}
-\frac{1}{2}\bar{\Sigma}_P\mathcal{S}^{\alpha\beta}\big)\Big)^{(1)}\frac{\sigma^{\mu\nu}}{2}.
\end{eqnarray}
The spin decomposition of the gain terms can be obtained similarly by taking the exchanges $\Sigma^>\leftrightarrow\Sigma^< $ and $S^<\leftrightarrow S^>$.

\end{appendix}

\bibliographystyle{iopart-num}
\bibliography{ref}

\providecommand{\newblock}{}
\begin{thebibliography}{10}
\expandafter\ifx\csname url\endcsname\relax
  \def\url#1{{\tt #1}}\fi
\expandafter\ifx\csname urlprefix\endcsname\relax\def\urlprefix{URL }\fi
\providecommand{\eprint}[2][]{\url{#2}}

\bibitem{Becattini:2013fla}
Becattini F, Chandra V, Del~Zanna L and Grossi E 2013 {\em Annals Phys.\/} {\bf
  338} 32--49 (\textit{Preprint} \eprint{1303.3431})

\bibitem{Fang:2016vpj}
Fang R~h, Pang L~g, Wang Q and Wang X~n 2016 {\em Phys. Rev.\/} {\bf C94}
  024904 (\textit{Preprint} \eprint{1604.04036})

\bibitem{Gao:2018jsi}
Gao J~h, Pang J~Y and Wang Q 2019 {\em Phys. Rev.\/} {\bf D100} 016008
  (\textit{Preprint} \eprint{1810.02028})

\bibitem{Liang:2004ph}
Liang Z~T and Wang X~N 2005 {\em Phys. Rev. Lett.\/} {\bf 94} 102301 [Erratum:
  Phys. Rev. Lett.96,039901(2006)] (\textit{Preprint} \eprint{nucl-th/0410079})

\bibitem{Voloshin:2004ha}
Voloshin S~A 2004  (\textit{Preprint} \eprint{nucl-th/0410089})

\bibitem{Betz:2007kg}
Betz B, Gyulassy M and Torrieri G 2007 {\em Phys. Rev.\/} {\bf C76} 044901
  (\textit{Preprint} \eprint{0708.0035})

\bibitem{Becattini:2007sr}
Becattini F, Piccinini F and Rizzo J 2008 {\em Phys. Rev.\/} {\bf C77} 024906
  (\textit{Preprint} \eprint{0711.1253})

\bibitem{STAR:2017ckg}
Adamczyk L {\em et~al.\/} (STAR) 2017 {\em Nature\/} {\bf 548} 62--65
  (\textit{Preprint} \eprint{1701.06657})

\bibitem{Adam:2018ivw}
Adam J {\em et~al.\/} (STAR) 2018 {\em Phys. Rev.\/} {\bf C98} 014910
  (\textit{Preprint} \eprint{1805.04400})

\bibitem{Acharya:2019vpe}
Acharya S {\em et~al.\/} (ALICE) 2020 {\em Phys. Rev. Lett.\/} {\bf 125} 012301
  (\textit{Preprint} \eprint{1910.14408})

\bibitem{Becattini:2013vja}
Becattini F, Csernai L and Wang D~J 2013 {\em Phys. Rev.\/} {\bf C88} 034905
  [Erratum: Phys. Rev.C93,no.6,069901(2016)] (\textit{Preprint}
  \eprint{1304.4427})

\bibitem{Becattini:2015ska}
Becattini F, Inghirami G, Rolando V, Beraudo A, Del~Zanna L, De~Pace A, Nardi
  M, Pagliara G and Chandra V 2015 {\em Eur. Phys. J.\/} {\bf C75} 406
  [Erratum: Eur. Phys. J.C78,no.5,354(2018)] (\textit{Preprint}
  \eprint{1501.04468})

\bibitem{Becattini:2016gvu}
Becattini F, Karpenko I, Lisa M, Upsal I and Voloshin S 2017 {\em Phys. Rev.\/}
  {\bf C95} 054902 (\textit{Preprint} \eprint{1610.02506})

\bibitem{Karpenko:2016jyx}
Karpenko I and Becattini F 2017 {\em Eur. Phys. J.\/} {\bf C77} 213
  (\textit{Preprint} \eprint{1610.04717})

\bibitem{Pang:2016igs}
Pang L~G, Petersen H, Wang Q and Wang X~N 2016 {\em Phys. Rev. Lett.\/} {\bf
  117} 192301 (\textit{Preprint} \eprint{1605.04024})

\bibitem{Xie:2017upb}
Xie Y, Wang D and Csernai L~P 2017 {\em Phys. Rev.\/} {\bf C95} 031901
  (\textit{Preprint} \eprint{1703.03770})

\bibitem{Becattini:2014yxa}
Becattini F, Bucciantini L, Grossi E and Tinti L 2015 {\em Eur. Phys. J. C\/}
  {\bf 75} 191 (\textit{Preprint} \eprint{1403.6265})

\bibitem{DeGroot:1980dk}
De~Groot S 1980 {\em {Relativistic Kinetic Theory. Principles and
  Applications}\/}

\bibitem{Kharzeev:2004ey}
Kharzeev D 2006 {\em Phys. Lett. B\/} {\bf 633} 260--264 (\textit{Preprint}
  \eprint{hep-ph/0406125})

\bibitem{Fukushima:2008xe}
Fukushima K, Kharzeev D~E and Warringa H~J 2008 {\em Phys. Rev. D\/} {\bf 78}
  074033 (\textit{Preprint} \eprint{0808.3382})

\bibitem{Neiman:2010zi}
Neiman Y and Oz Y 2011 {\em JHEP\/} {\bf 03} 023 (\textit{Preprint}
  \eprint{1011.5107})

\bibitem{Son:2012bg}
Son D and Spivak B 2013 {\em Phys. Rev. B\/} {\bf 88} 104412 (\textit{Preprint}
  \eprint{1206.1627})

\bibitem{Son:2012wh}
Son D~T and Yamamoto N 2012 {\em Phys. Rev. Lett.\/} {\bf 109} 181602
  (\textit{Preprint} \eprint{1203.2697})

\bibitem{Son:2012zy}
Son D~T and Yamamoto N 2013 {\em Phys. Rev. D\/} {\bf 87} 085016
  (\textit{Preprint} \eprint{1210.8158})

\bibitem{Stephanov:2012ki}
Stephanov M and Yin Y 2012 {\em Phys. Rev. Lett.\/} {\bf 109} 162001
  (\textit{Preprint} \eprint{1207.0747})

\bibitem{Pu:2010as}
Pu S, Gao J~h and Wang Q 2011 {\em Phys. Rev. D\/} {\bf 83} 094017
  (\textit{Preprint} \eprint{1008.2418})

\bibitem{Chen:2012ca}
Chen J~W, Pu S, Wang Q and Wang X~N 2013 {\em Phys. Rev. Lett.\/} {\bf 110}
  262301 (\textit{Preprint} \eprint{1210.8312})

\bibitem{Hidaka:2016yjf}
Hidaka Y, Pu S and Yang D~L 2017 {\em Phys. Rev. D\/} {\bf 95} 091901
  (\textit{Preprint} \eprint{1612.04630})

\bibitem{Huang:2018wdl}
Huang A, Shi S, Jiang Y, Liao J and Zhuang P 2018 {\em Phys. Rev. D\/} {\bf 98}
  036010 (\textit{Preprint} \eprint{1801.03640})

\bibitem{Liu:2018xip}
Liu Y~C, Gao L~L, Mameda K and Huang X~G 2019 {\em Phys. Rev. D\/} {\bf 99}
  085014 (\textit{Preprint} \eprint{1812.10127})

\bibitem{Lin:2019ytz}
Lin S and Shukla A 2019 {\em JHEP\/} {\bf 06} 060 (\textit{Preprint}
  \eprint{1901.01528})

\bibitem{Hattori:2019ahi}
Hattori K, Hidaka Y and Yang D~L 2019 {\em Phys. Rev. D\/} {\bf 100} 096011
  (\textit{Preprint} \eprint{1903.01653})

\bibitem{Wang:2019moi}
Wang Z, Guo X, Shi S and Zhuang P 2019 {\em Phys. Rev. D\/} {\bf 100} 014015
  (\textit{Preprint} \eprint{1903.03461})

\bibitem{Gao:2019znl}
Gao J~H and Liang Z~T 2019 {\em Phys. Rev. D\/} {\bf 100} 056021
  (\textit{Preprint} \eprint{1902.06510})

\bibitem{Weickgenannt:2019dks}
Weickgenannt N, Sheng X~L, Speranza E, Wang Q and Rischke D~H 2019 {\em Phys.
  Rev. D\/} {\bf 100} 056018 (\textit{Preprint} \eprint{1902.06513})

\bibitem{Liu:2020flb}
Liu Y~C, Mameda K and Huang X~G 2020 {\em Chin. Phys. C\/} {\bf 44} 094101
  (\textit{Preprint} \eprint{2002.03753})

\bibitem{Yang:2020hri}
Yang D~L, Hattori K and Hidaka Y 2020 {\em JHEP\/} {\bf 20} 070
  (\textit{Preprint} \eprint{2002.02612})

\bibitem{Weickgenannt:2020aaf}
Weickgenannt N, Speranza E, Sheng X~l, Wang Q and Rischke D~H 2020
  (\textit{Preprint} \eprint{2005.01506})

\bibitem{Carignano:2019zsh}
Carignano S, Manuel C and Torres-Rincon J~M 2020 {\em Phys. Rev. D\/} {\bf 102}
  016003 (\textit{Preprint} \eprint{1908.00561})

\bibitem{Li:2019qkf}
Li S and Yee H~U 2019 {\em Phys. Rev. D\/} {\bf 100} 056022 (\textit{Preprint}
  \eprint{1905.10463})

\bibitem{Hou:2020mqp}
Hou D and Lin S 2020  (\textit{Preprint} \eprint{2008.03862})

\bibitem{Klevansky:1997wm}
Klevansky S~P, Ogura A and Hufner J 1997 {\em Annals Phys.\/} {\bf 261} 37--73
  (\textit{Preprint} \eprint{hep-ph/9708263})

\bibitem{Blaizot:2001nr}
Blaizot J~P and Iancu E 2002 {\em Phys. Rept.\/} {\bf 359} 355--528
  (\textit{Preprint} \eprint{hep-ph/0101103})

\bibitem{Florkowski:2019gio}
Florkowski W, Kumar A and Ryblewski R 2020 {\em Acta Phys. Polon.\/} {\bf B51}
  945--959 (\textit{Preprint} \eprint{1907.09835})

\end{thebibliography}

\end{document}